# Unusual dissociative adsorption of $H_2$ over stoichiometric MgO thin film supported on molybdenum

Zhenjun Song*, Hu Xu

Department of Physics, South University of Science and Technology of China, Shenzhen, 518055, China

Corresponding author: Zhenjun Song

Address: Department of Physics, South University of Science and Technology of China, Shenzhen, 518055, China

E-mail: songzj@mail.sustc.edu.cn

Tel: +86 13516213950

**Abstract**

   The dissociation of a hydrogen molecule on MgO(001) films deposited on Mo(001) surface is investigated systematically using periodic density-functional theory method. The unusual adsorption behavior of heterolytic dissociative hydrogen molecule at neighboring surface oxygen and surface magnesium, is clarified here. To my knowledge, this heterolytic dissociative state has never been found before on bulk MgO(001) or metal supported MgO(001) surfaces. The results confirm that, in all cases, the heterolytic dissociation is much more favorable that homolytic dissociation both energetically and kinetically. The energy difference between two dissociative states are very large, in the range of 1.1 eV ~ 1.5 eV for Mo supported 1 ML ~ 3 ML oxide films, which inhibits, to a great extent, the homolytic dissociation in the respect of reaction thermodynamics. The energy barrier of heterolytic dissociation are about 0.5 eV, much lower that the barrier of homolytic dissociation. The transformation reaction on thick films will be more endothermic. Passing through heterolytic dissociation state have significantly lowered the reaction heat and the energy barrier for obtaining homolytic dissociative structure, which make the homolytic splitting of $H_2$ easier on 2 ML oxide films. The results provides a useful strategy for enhancing the reactivity of the nonreducible metal oxide.

# 1. Introduction

Dihydrogen ($H_2$), a carrier of most promising green energy, can be converted to thermal energy without producing poisonous substances and by product. Moreover, the rates and efficiencies of renewable production of hydrogen by light – driven water splitting continue to increase[1-5], which has been considered as an ultimate solution to energy and environmental issues. Hydrogen adsorption and diffusion is fundamentally important to physical and chemical elementary processes on surfaces[6]. In the presence of hydrogen, the coupling of hydrogen with CO, followed by C–O bond cleavage is found on Ru(0001) surface, which lowers the C–O fragmentation barrier[7-10]. The hydrogenation process on surfaces, which is responsible for the enhanced interaction with the surface, is predicted to be important for this hydrogen-assisted dissociation mechanism. The alkyne hydrogenation catalyst Pd is highly permeable for hydrogen[11]. In which way the indispensable hydrogen atoms absorbed beneath the top layer influence surface catalytic reactivity is controversially debated[12]. The catalytic effect of single crystal is less active than nanoclusters, and the pure nanoclusters of a similar size accurately mimic the nanoclusters deposited on inert metal oxides[12-15]. CO oxidation, as an environmentally important topic in heterogeneous catalysis, have enhanced catalytic activity on $Ag/CeO_2$ due to the presence of scattered $Ag^+$ ions[16]. Recent studies on oxygen activation and CO oxidation reveal that, compared with clusters with even number of electrons, clusters with an odd number of electrons shown enhanced catalytic ability and efficiency[17-19]. Interestingly, the hydrogen atoms can regulate electron parity to satisfy the catalytic requirement and improve the reactivity[17, 20, 21]. The hydrogen atoms can form by hydrogen adsorption and dissociation on substance surfaces.

Although many studies pay attention to the adsorption behavior of hydrogen on metal surfaces and metal clusters, much less research work investigate the adsorption and decomposition of hydrogen gas on metal oxides, especially insulators, because of its chemical inertness. The α–$Ga_2O_3$, activated under vacuum at 773 K, is able to dissociate hydrogen molecule into $H^+$ and $H^-$ species, leading to the formation of Ga–H species[22]. The behavior of hydrogen molecules toward heterolytic dissociative chemisorption, on microcrystalline MgO, CaO, and SrO is studied, using UV−vis diffuse relectance spectroscopy[23]. Generally, on nonreducible oxides, $H_2$ dissociates heterolytically with formation of a proton adsorbed on the oxide $O^{2-}$ anions and a hydride anion adsorbed on the surface cations. Heterolytic dissociation of $H_2$ on MgO powder has been found experimentally[23-25] and confirmed theoretically by ab initio calculations[26, 27]. MgO is known to catalyze various reactions containing hydrocarbons[26], such as, $H_2/D_2$ exchange reaction[28, 29], methanol and ethanol decomposition[30-33], alkene hydrogenation[34], low temperature water gas shift reaction[35, 36], et al, where the atomic and molecular hydrogen adsorption properties on MgO surfaces may influence the reactions greatly. In this work, I perform a systematic investigation on the hydrogen adsorption properties on molybdenum supported stoichiometric MgO(001) surfaces. The results reveal novel dissociative adsorption behaviors and new adsorption states whose energy barriers are extraordinarily low. To my knowledge, the heterolytic dissociative state, which has been clarified here, has never been found before on bulk MgO(001) or metal supported MgO(001) surfaces. The successful attempts to elucidate the abnormal adsorption and splitting properties of hydrogen on molybdenum supported stoichiometric MgO(001) surfaces may aid in our completely new understanding of the reactivity of the nonreducible metal oxide.

## 2. Methods

The spin-polarized density functional calculations have been performed by using Vienna *ab initio* simulation package (VASP)[37, 38] to determine all structural, energetic and electronic results. I choose Perdew-Burke-Ernzerhof (PBE) functional[39] within generalized gradient approximation (GGA) to describe exchange and correlation effects, which includes an accurate description of the uniform electron gas, correct behavior under uniform scaling and a smoother potential compared with Perdew-Wang-

1991 (PW91) functional. Projector Augmented Wave (PAW)[37, 40] technique is used to describe electronic structure and treat the interactions between valence electrons and the core. Electron configurations $1s^1$, $2s^22p^4$, $2s^2$, $4d^55s^1$ are used to describe valence electrons in H, O, Mg and Mo atoms. The Kohn-Sham orbitals were expanded by using plane waves with a kinetic energy cutoff of 500 eV. Through spin-polarized plane wave calculations with a k-point mesh of $9\times9\times9$, the lattice constants of the Mo bulk are determined to be 3.151 Å, which are in good agreement to reported experimental value 3.14 Å[41]. I use a four atomic layer Mo slab, with the two bottom layers fixed at bulk positions while the other two metal layers and the MgO film are fully relaxed until all atomic Hellmann-Feynman forces are less than 0.02 eV/Å. I found that an even larger number of Mo layers do not change the surface chemical properties of the MgO films. Convergence criterion for energy minimization is $1.0\times10^{-5}$ eV. For the calculations, we use supercells containing 16Mg + 16O atoms per layer or 16Mo atoms per layer. Gamma-centered k-point meshes $2\times2\times1$ and $4\times4\times1$ is used to sample the first Brillouin zone, for structure optimization and energy calculation respectively.

In all calculations the periodically repeated slabs are separated by a thick vacuum larger than 17 Å. The energy barriers and transition states are located by using the climbing image nudged elastic band (CI-NEB) method[42], which is an efficient method for searching the minimum energy path (MEP) connecting the given initial and final states. Because the highest-energy image is trying to maximize its energy along the band and minimize energy in all other directions, the exact saddle point along the reaction path is easier to find. Therefore, less number of intermediate images is needed in CI-NEB than NEB.

## 3. Results and discussion

For hydrogen molecular adsorption on the bare MgO(001) surface and MgO/Mo(001) surfaces, due to the weak $H_2\cdots O_s$ and $H_2\cdots Mg_s$ interactions, the adsorption energies are very small, and the $H_2$ molecules can translate over the surface easily. Diverse dissociative adsorption configurations, including classic heterolytic conformations and homolytic conformations are examined. For the heterolytic dissociation pathway, the hydrogen molecule attacks surface oxygen and the neighboring magnesium atom, leading to the breaking of H–H covalent bond and the formation of $O_s$–H and $Mg_s$–H bonds. With regard to homolytic dissociation pathway, the hydrogen molecule attacks two diagonally adjacent surface oxygen atoms, which result in the complete splitting of H–H covalent bond and the formation of two new strong $O_s$–H covalent bonds.

The heterolytic dissociative state, with two hydrogen atoms adsorbed on neighboring O and Mg atoms cannot form after relaxation on bare MgO(001). The heterolytic dissociative state which exists on bare MgO(001) with favored total energy, have an adsorption structure with two hydrogen atoms bonded to one surface oxygen atom, and one next nearest surface magnesium atom 4.82 Å far away from that surface oxygen atom. However, in our calculation this heterolytic dissociative structure is 1.84 eV higher in energy than the molecular adsorption state. Thus, the formation of heterolytic dissociative structure of $H_2$ is very difficult on bare MgO(001) surface. Furthermore, the homolytic dissociative structure of $H_2$ is 3.06 eV higher in energy than the molecular adsorption state, which is nearly impossible to realize on bare MgO(001) surface, under normal condition.

Table 1. Calculated heterolytically dissociative adsorption energies of $H_2$ ($\Delta E_{het}$), homolytically dissociative adsorption energies of $H_2$ ($\Delta E_{hom}$), energy barriers of dissociation ($E_a$(het), and $E_a$(hom)), and reaction energies and energy barriers of transformation from heterolytic dissociative state to homolytic dissociative state ($\Delta E_{tr}$ and $E_a$(tr)). The unit of energy is in eV.

| Surfaces | $\Delta E_{het}$ | $E_a$(het) | $\Delta E_{hom}$ | $E_{a1}$(hom) | $E_{a2}$(hom) | $\Delta E_{tr}$ | $E_a$(tr) |
|---|---|---|---|---|---|---|---|
| 1 ML MgO/Mo(001) | 0.48 | 0.50 | 0.79 | 0.64 | 0.97 | 0.31 | 1.06 |
| 2 ML MgO/Mo(001) | 0.39 | 0.48 | 1.45 | 1.86 | —— | 1.06 | 1.44 |
| 3 ML MgO/Mo(001) | 0.48 | 0.54 | 1.66 | 0.72 | 1.28 | 1.17 | 1.46 |

To facilitate the dissociation adsorption of hydrogen molecule on MgO(001) surface with low catalytic activity, transition metal supported oxide films (MgO/Mo(001)) has been considered. Calculated heterolytic dissociative adsorption energies of $H_2$ ($\Delta E_{het}$), homolytic dissociative adsorption energies of $H_2$ ($\Delta E_{hom}$), energy barriers of dissociation ($E_a$(het), and $E_a$(hom)), and reaction energies and energy barriers of transformation from heterolytic dissociative state to homolytic dissociative state ($\Delta E_{tr}$ and $E_a$(tr)) are summarized in Table 1. Relevant optimized geometrical parameters obtained using PBE functional at heterolytic dissociative state and homolytic dissociative state are listed in Table 2. The homolytic and heterolytic bond dissociation mechanisms are searched by CI–NEB method. The minimum energy pathway (MEP) connects two local minima, namely, the molecular hydrogen ca. 4 Å far away from surface, which interacts very weakly with the surface, as the initial state, and the dissociative adsorption structure as the final state. Important structures of reactants and products at the extreme points of MEP are displayed in Figure 1. The converged CI–NEB results for heterolytic dissociation and homolytic dissociation, together with the transformation pathway from heterolytic dissociation states to homolytic dissociation states are displayed in Figure 2.

Table 2. Relevant optimized geometrical parameters at heterolytic dissociative state and homolytic dissociative state.

| Bond length (Å) | Heterolytic | | | Homolytic | | |
| --- | --- | --- | --- | --- | --- | --- |
| | 1 ML | 2 ML | 3 ML | 1 ML | 2 ML | 3 ML |
| H1–H2 | 1.20 | 1.29 | 1.25 | 3.00 | 2.58 | 2.56 |
| H1–O1 | 1.09 | 1.06 | 1.08 | 0.98 | 1.01 | 1.01 |
| H2–Mg1 | 1.86 | 1.86 | 1.87 | 2.72 | 2.47 | 1.87 |
| H2–O2 | 2.98 | 3.01 | 2.99 | 0.98 | 1.01 | 1.01 |
| O1–Mg1 | 2.87 | 2.79 | 2.77 | 3.24 | 2.87 | 2.86 |
| O1–Mg2 | 2.33 | 2.39 | 2.38 | 3.25 | 2.85 | 2.85 |
| O2–Mg1 | 2.33 | 2.36 | 2.35 | 3.26 | 2.87 | 2.85 |
| O2–Mg2 | 2.49 | 2.29 | 2.26 | 3.25 | 2.86 | 2.85 |

As shown in Table 2, the calculated bond distances of H1–H2 are 1.20 Å, 1.29 Å, 1.25 Å for 1 ML ~ 3 ML oxide films, indicating the successful heterolytic dissociation of $H_2$ (bond length, 0.75 Å). In the contrast, at the homolytic dissociative state, the distances of H1–H2 are more large, namely, 3.00 Å, 2.58 Å and 2.56 Å for 1 ML ~ 3 ML oxide films, betokening the more thorough breaking of H–H covalent bond. At heterolytic dissociative state, the bond length of H1–O1 are 1.09 Å, 1.06 Å and 1.08 Å, and the bond length of H2–Mg1 are 1.86 Å, 1.86 Å and 1.87 Å for 1ML ~ 3 ML oxide films, which verify the heterolytic dissociation of $H_2$ and the formation of new hydroxyl group and H–Mg ionic bond. At the homolytic dissociative state, the H1–O1 bond distances are 0.98 Å, 1.01 Å, and 1.01 Å for 1 ML ~ 3 ML oxide films. The H1–O1 and

H2–O2 are exactly the same, demonstrating the homolytic splitting of $H_2$ and the formation of two identical O–H bonds. The surface chemical bonds at the reaction site are also examined. The results show that for heterolytic dissociation, the surface structure deformation is less severe, compared with that of homolytic dissociation. At the homolytic dissociative state, the bond length of surface O–Mg are larger than 2.85 Å. For 1 ML oxide film, especially, the surface O–Mg bond are longer than 3.25 Å for 1 ML oxide film. Thus, the homolytic dissociation will distort the surface ionic bonding more significantly, which may be related to the high reaction heat absorption, and very high energy barrier of homolytic dissociation. On the contrast, for heterolytic dissociation, the O–Mg bonds are shorter than 2.49 Å except the O1–Mg1 (2.77 Å ~ 2.87 Å) which is directly connected with hydrogen atoms. At the heterolytic dissociative state, the change of surface thickness affect the surface bonding slightly (< 0.2 Å). At the homolytic dissociative state, the surface bonding are changed more greatly, by ca. 0.4 Å from 2 ML to 1 ML oxide film. Therefore, the oxide film thickness should affect the homolytic dissociation greatly, as verified by the surface structure distortion.

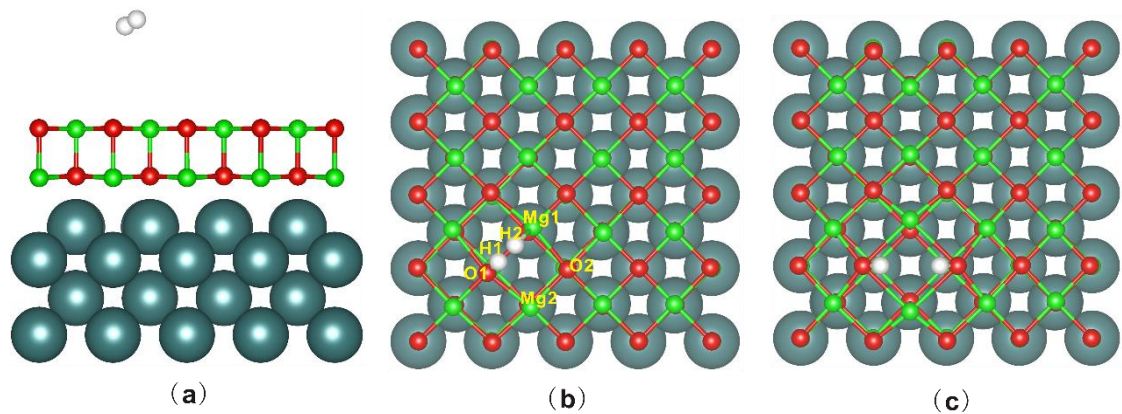

(a)　　　　　　　　　(b)　　　　　　　　　(c)

Figure 1. (a). Optimized structure of hydrogen molecule interacting with Mo supported 2 ML MgO(001) film very weakly, (b). Optimized structure of heterolytic dissociative hydrogen molecule adsorbing on Mo supported 2 ML MgO(001) film, (c). Optimized structure of homolytic dissociative hydrogen molecule adsorbing on Mo supported 2 ML MgO(001) film.

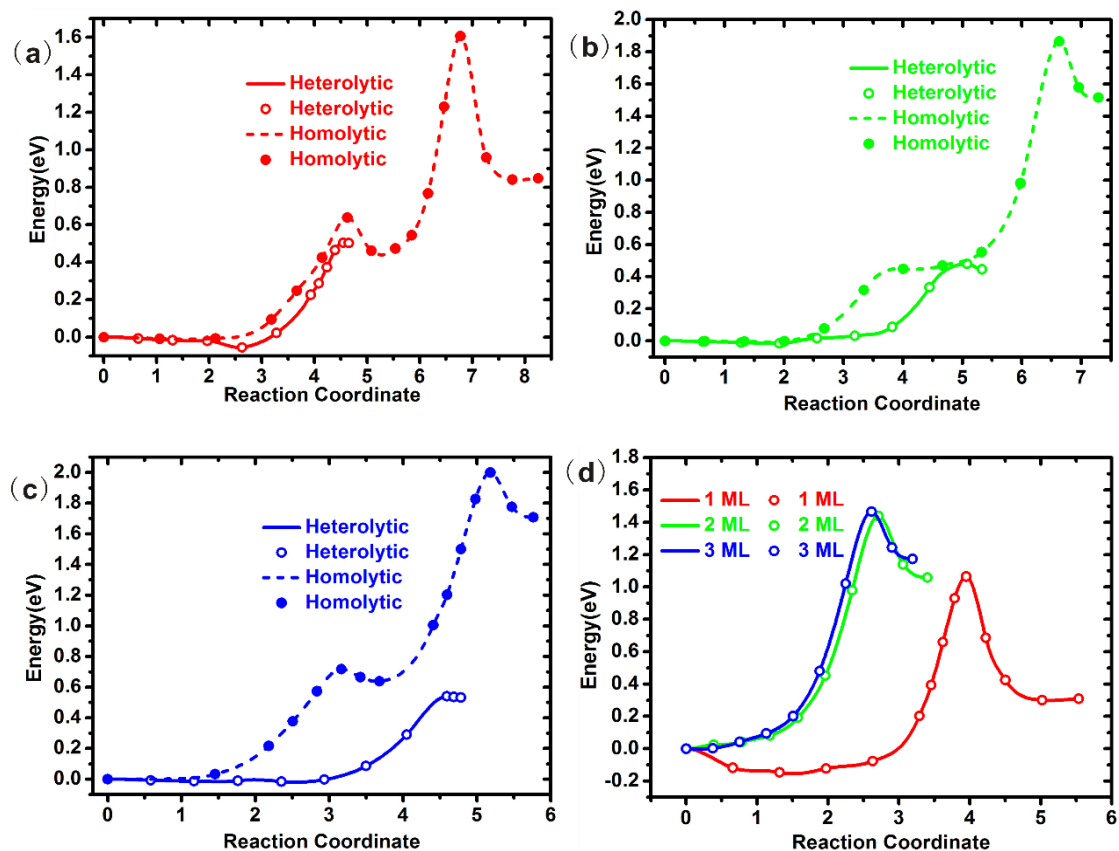

Figure 2. (a). Reaction pathway for heterolytic and homolytic dissociation of hydrogen molecule on Mo supported 1 ML MgO(001) film, (b). Reaction pathway for heterolytic and homolytic dissociation of hydrogen molecule on Mo supported 2 ML MgO(001) film, (c). Reaction pathway for heterolytic and homolytic dissociation of hydrogen molecule on Mo supported 3 ML MgO(001) film, (d). Transformation pathway from heterolytically dissociative state to homolytically dissociative state of hydrogen molecule on Mo supported 1 ~ 3 ML MgO(001) films.

The results confirm that, in all cases, the heterolytic dissociation is much more favorable than homolytic dissociation, both energetically and kinetically. The energy difference between two dissociative states are very large, in the range of 1.1 eV ~ 1.5 eV for 1 ML ~ 3 ML oxide films, which inhibits, to a great extent, the homolytic dissociation on the MgO/Mo(001) surfaces, in the respect of reaction thermodynamics. According to the activation energies presented in Table 1, the energy barriers of heterolytic dissociation are 0.50 eV, 0.48 eV, and 0.54 eV for 1 ML ~ 3 ML oxide films. However, the energy barriers to cleave H–H covalent bond homolytically increase by more than two times to 1.86 eV for 2 ML oxide films. The reactions occurring on odd-layers oxide films must experience two barriers, 0.64 eV, 0.97 eV for 1 ML oxide film and 0.72 eV, 1.28 eV for 3 ML oxide film. From the $\Delta E_{\text{hom}}$ and $E_a(\text{hom})$ shown in Table 1, it can be clearly inferred that comparing with 1 ML oxide film the thicker films are extremely unfavorable for homolytic dissociation of $H_2$. However, the heterolytic dissociation do not show large difference in energy barrier, for reactions taking place on 1 ML ~ 3 ML oxide films.

Because the heterolytic dissociation with small reaction heat absorption and low energy barrier is more favorable, the assumption can be boldly made that the homolytic dissociation state can be obtained by splitting $H_2$ heterolytically, and then transforming heterolytic dissociation state to homolytic dissociation state. In this way the energy barrier of homolytic dissociation on 2 ML oxide film can be probably lowered. Thus the investigation of the transformation between two dissociative states is of important significance. From the transformation pathway (Figure 2d) and energetic results (Table 1), it can be found that from heterolytically dissociative state to homolytically dissociative state of $H_2$, all the researched systems are endothermic. The reaction on 1 ML oxide film absorbs least heat, and the reactions on thicker films will be more endothermic. The energy barrier of transformation show similar energy sequence. Transformation reactions on thinner films have lower energy barriers. The transformation reaction on 1 ML oxide film is only slightly endothermic by 0.31 eV, while the energy barrier is relatively high, 1.06 eV. The energy barrier for transformation on 1 ML and 3 ML is larger than that of homolytic splitting $H_2$ directly. It is worth mentioned that, passing through heterolytic dissociative state has significantly lowered the reaction heat (1.45 eV) and the energy barrier (1.86 eV) for obtaining homolytic dissociative structure, which make the homolytic splitting of $H_2$ on 2 ML oxide film easier.

Table 3. The calculated effective Bader charges (unit in electron) on H1, H2, O1, O2, Mg1, Mg2 and Mo atoms, at heterolytic dissociative state and homolytic dissociative state.

| Bader charge | Heterolytic | | | Homolytic | | |
|---|---|---|---|---|---|---|
| | 1 ML | 2 ML | 3 ML | 1 ML | 2 ML | 3 ML |
| H1 | +0.47 | +0.50 | +0.49 | +0.48 | +0.28 | +0.27 |
| H2 | −0.60 | −0.65 | −0.63 | +0.49 | +0.29 | +0.26 |
| O1 | −1.46 | −1.52 | −1.53 | −1.44 | −1.56 | −1.56 |
| O2 | −1.48 | −1.65 | −1.65 | −1.45 | −1.56 | −1.56 |
| Mg1 | +1.61 | +1.64 | +1.64 | +1.40 | +1.41 | +1.42 |
| Mg2 | +1.64 | +1.67 | +1.67 | +1.40 | +1.42 | +1.42 |
| Mo | −1.88 | −1.51 | −1.44 | −2.54 | −1.50 | −1.44 |

To understand the influence of hydrogen adsorption and dissociation on the charge distribution of Mo supported MgO(001) film and the adsorbates, effective Bader charge analysis is conducted and shown in Table 3. At heterolytic dissociative state, the hydrogen atoms connected with surface oxygen show positive charges, and the hydrogen atoms attached to surface magnesium show negative charges. It is worth noted that the H1 and H2 atoms adsorbing on 2 ML oxide film, carry more charges than that adsorbing on 1 ML and 3 ML oxide films, indicating the strongest interaction between 2 ML oxide film and H atoms. O1 atom carries less negative charges than O2, due to the formation of covalent bond between H1 and O1. Because of the formation of the H–Mg bond with partial covalent character, Mg1 atom carries less positive charges than Mg2. The amount of charge transfer of Mo is larger for thinner films (−1.88 e, −1.51 e, and −1.44 e for Mo supported 1 ML ~ 3 ML oxide films respectively).

At the homolytic dissociative state, the H1 and H2 atoms lose electrons and carry almost the same amount of positive charges, indicating the bonding environment of H1 and H2 resemble each other. As the film thickness increases, the amount of positive charges of H1 and H2 decrease sharply. Comparing with Mo supported thicker films, on the Mo supported 1 ML MgO(001), the H atoms adsorb on the surface oxygen more firmly, which is due to the unique structural feature of 1 ML MgO. The Mo supported 1 ML MgO(001) have lower coordinated (tetra–coordinated) oxygen atoms, while the surface oxygen atoms of Mo supported thicker films are all penta-coordinated. The O1 and O2 atoms are negatively charged at the same level. As the thickness increases, O1 and O2 are more negatively charged, indicating the strengthening of the ionicity. The charge amount of Mg1 and Mg2 atoms at homolytic dissociative state are significantly less than that of Mg1 and Mg2 atoms at heterolytic dissociative state. This fact suggests that the homolytic dissociation of $H_2$ and the formation of strong covalent bonds $O_s$–H leads to the insufficient bonding of surface magnesium atoms located near the reaction site. At homolytic dissociative state, the Mo substrate of supported 1 ML oxide film obtain much more electrons than that at heterolytic dissociative state. As the hydrogen atoms all donate electrons to the surface, the charge transfer behavior is very different at two dissociative states.

The differential charge density contour for heterolytic dissociative $H_2$ which can be obtained with very low energy barrier (0.48 eV) is shown in Figure 3. The differential charge density is obtained from $\rho = \rho(\text{total}) - \rho(\text{H1+H2}) - \rho(\text{MgO}) - \rho(\text{Mo})$, where $\rho(\text{total})$, $\rho(\text{H1+H2})$, $\rho(\text{MgO})$, and $\rho(\text{Mo})$ denote the charge densities of the optimized whole system, the adsorbed hydrogen atoms, MgO films, and Mo substrates. The charge accumulation occurs for the hydrogen connected to surface magnesium. The charge depletion occurs for the hydrogen bonded to surface oxygen. Charge transfer between interface Mo and O atoms can be clearly seen. Thus the covalent bonding between Mo and O definitely formed. Giant charge density accumulation take place at the interfacial region, demonstrating unambiguously the effective interaction between interfacial Mo and O atoms. The density of states for H1, H2, $O_s$, $Mg_s$, interface $O_{2p}$ and interface $Mo_{4d}$ are calculated and illustrated in Figure 4. The interface $O_{2p}$ and interface $Mo_{4d}$ show large amount of overlapping area, verifying again the strong Mo–O interaction and formation of covalent bond. The H1 bonded to surface oxygen, show new electronic state peak at around −8.5 eV. H2 atom bonded to surface magnesium, show new state peak at around −2.5 eV. These electron distribution features, which can not be observed in pure MgO, may play an essential role in enhancing the reactivity of MgO toward splitting hydrogen molecule.

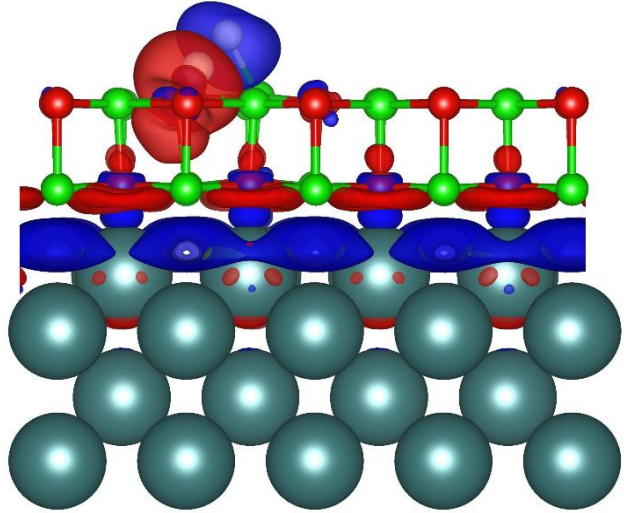

Figure 3. The differential charge density contour in unit of e/bohr$^3$ calculated by subtracting from charge difference between the whole system and the adsorbed hydrogen atoms, the Mo supported 2 ML MgO(001), which is defined as $\rho = \rho(total) - \rho(H1+H2) - \rho(MgO) - \rho(Mo)$. The contour is plotted with charge density value of $\pm 0.0025$ e/bohr$^3$. The blue color is density gain and the red color is density loss.

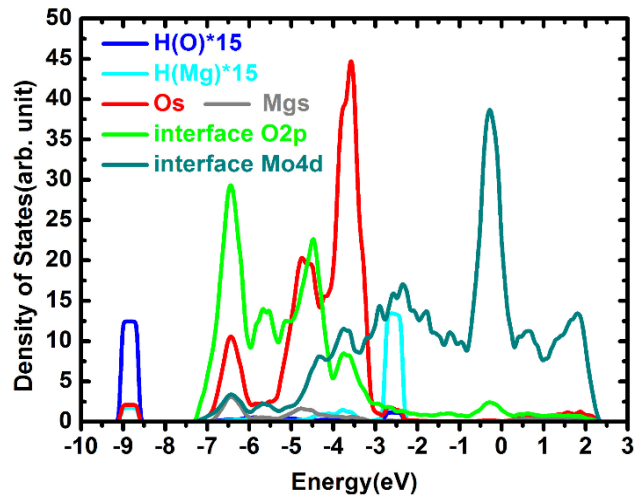

Figure 4. Density of states of H1, H2, $O_s$, $Mg_s$, interface $O_{2p}$, and interface $Mo_{4d}$.

## 4. Conclusion

In conclusion, the hydrogen molecule dissociating heterolytically and homolytically on MgO(001) films deposited on Mo(001) surface is investigated using density-functional theory method. The results confirm that, in all cases, the heterolytic dissociation is much more favorable that homolytic dissociation both energetically and kinetically. The energy difference between two dissociative states are very large, in the range of 1.1 eV ~ 1.5 eV for Mo supported 1 ML ~ 3 ML oxide films, which inhibits, to a great extent, the homolytic dissociation in the respect of reaction thermodynamics. The energy barrier of heterolytic dissociation are about 0.5 eV, much lower that the barrier of homolytic dissociation. The behavior of the relatively nonreactive MgO toward splitting hydrogen molecule is enhanced dramatically (reaction heat are reduced by many times) by supporting on the Mo substrate.

**Acknowledgements**

This work was financially supported by National Natural Science Foundation of China (NSFC Grant No.11204185 and 11334003). The author also acknowledge the National Supercomputing Center in Shenzhen for providing the computational resource.